\documentclass[conference]{IEEEtran}
\IEEEoverridecommandlockouts
\usepackage{cite}
\usepackage{amsmath,amssymb,amsfonts}
\usepackage{algorithmic}
\usepackage{graphicx}
\usepackage{textcomp}
\usepackage{xcolor}
\def\BibTeX{{\rm B\kern-.05em{\sc i\kern-.025em b}\kern-.08em
    T\kern-.1667em\lower.7ex\hbox{E}\kern-.125emX}}
\begin{document}

\title{Embedding technique and network analysis of scientific innovations emergence in an
arXiv-based concept network}

\author{\IEEEauthorblockN{Serhii Brodiuk}
\IEEEauthorblockA{\textit{Faculty of Applied Sciences, Ukrainian Catholic University}\\
UA--79011 Lviv, Ukraine \\
brodiuk@ucu.edu.ua}
\and
\IEEEauthorblockN{Vasyl Palchykov}
\IEEEauthorblockA{\textit{$\mathbb{L}^4$ Collaboration \& Doctoral College} \\
\textit{for the Statistical Physics of Complex Systems}\\
Leipzig-Lorraine-Lviv-Coventry, Europe \\
palchykov@gmail.com}
\and
\IEEEauthorblockN{Yurij Holovatch}
\IEEEauthorblockA{\textit{Institute for Condensed Matter Physics, National Acad. Sci.of Ukraine}\\
UA--79011 Lviv, Ukraine \\
\textit{$\mathbb{L}^4$ Collaboration \& Doctoral College for the Statistical Physics of Complex Systems}\\
Leipzig-Lorraine-Lviv-Coventry, Europe \\
\textit{Centre for Fluid and Complex Systems, Coventry University}\\
Coventry, CV1 5FB, United Kingdom \\
hol@icmp.lviv.ua}
}

\maketitle

\begin{abstract}

Novelty is an inherent part of innovations and discoveries.
Such processes may be considered as an appearance of new ideas or as an emergence of atypical connections between the 
existing ones. The importance of such connections hints for investigation of innovations through network or 
graph representation in the space of ideas. In such representation, a graph node corresponds to the relevant 
concept (idea), whereas an edge between two nodes means that the corresponding concepts have been used in a common context. 
In this study we address the question about a possibility to identify the edges between existing concepts 
where the innovations may emerge. 
To this end, we use a well-documented scientific knowledge landscape of 1.2M \texttt{arXiv}.org manuscripts 
dated starting from April 2007 and until September 2019. We extract relevant concepts for them using the 
ScienceWISE.info platform.
Combining approaches developed in complex networks science and graph embedding, we discuss the predictability 
of edges (links) on the scientific knowledge landscape where the innovations may appear.
\end{abstract}


\section{Introduction}

An idea of scientific analysis of science is not new.
It is at least as old as the science itself, see, e. g. \cite{zhmud2008origin} and references therein.
Contemporary studies in this domain share a common specific feature: besides traditional philosophical and culturological context, such analysis attains quantitative character.
The questions of interest cover a wide spectrum, ranging from fundamental, such as: what is the structure of science? How do its constituents interact?
How does knowledge propagate? \cite{lewens2016meaning,mryglod2016quantifying,Krenn20} to entirely practical ones: which fields of science 
deserve financial  investments or how to rate scientists in a particular domain? \cite{leydesdorff2015scientometrics,berche2016academic,Thurner19}.
All these and many more questions constitute a subject of a science of science or logology \cite{zeng2017science}.

The problem we consider in this paper concerns an emergence of new scientific knowledge or the so-called scientific innovation.
Quantitative investigation and modeling of innovations are not straightforward.
On the one hand, one may think of innovation as an emergence of a new idea, see, e. g. \cite{iacopini2018network}.
Another approach considers innovation as an atypical combination of existing ideas, see, e. g. \cite{uzzi2013atypical}.
The goal of our work is to suggest a way to quantify analysis of scientific innovations emergence
and to propose an approach to identify edges on the graph of knowledge where innovations may emerge. 
We believe that such analysis, if successful, is useful both from the fundamental point of view, explaining properties 
of knowledge formation, as well as is of practical relevance, helping to detect innovation-rich fields. 

To reach this goal, we will analyze a body of scientific publications taking an \texttt{arXiv} repository of research papers 
\cite{arxiv} and studying its dynamics with a span of time.
We will use a specially tailored software, ScienceWISE.info platform \cite{sciencewise}, to extract a set of concepts from 
all publications on an annual basis.
These are the properties of this set of concepts that will serve us as a proxy of structural features and dynamics of human knowledge.
In particular, we will use complex network theory
\cite{holovatch2006complex,newman2010networks,barabasi2016network,holovatch2018statistical}
to track intrinsic connections between concepts that are contained in different papers.
Using several completing each other approaches we will construct a complex network of concepts (as a proxy of a complex network of knowledge)
and we will calculate its main topological characteristics, paying particular attention to the emergence of new links between existing concepts.
These last may serve as a signal about an emergence of atypical combinations between existing ideas, i. e. about scientific innovations.
We will refine our analysis by exploiting embedding technique \cite{mikolov2013distributed,lerer2019pytorch} to quantify a proximity 
measure between different concepts and in this way, we will establish a solid and falsifiable procedure to quantify an emergence of possible 
scientific innovations in certain fields of science.

The rest of the paper is organized as follows.
In the next Section \ref{II} we describe the dataset used in the analysis, the data is presented 
in complex network form and analysed in Section \ref{III}.
In Section \ref{IV} we introduce the concept embedding technique and study dynamics 
of link appearance. The results are summarized in the last Section \ref{V}.

\section{Dataset} \label{II}
We use the E-repository of preprints \texttt{arXiv}.org \cite{arxiv} as a source of data:
at the moment of writing this paper, there are about  1.6M full-text accesible manuscripts 
uploaded to the \texttt{arXiv}. It makes them an optimal source
to extract scientific ideas/concepts. The \texttt{arXiv} covers a variety of scientific fields 
such as physics, mathematics, computer science, quantitative biology, quantitative finance, 
statistics, electrical engineering and systems science, and economics.
The average daily upload rate is 400 $\approx$ 12.5K new manuscripts per month (every next year 
there are $\approx$ 5000 more articles than the previous one, starting from 1991).
Each paper submitted to the \texttt{arXiv} contains, besides the full-text, different metadata 
such as authors, subject category (categories), journal reference, DOI 
 if any, submissions history with dates, etc.
For the purpose of our study, we need to extract specific words or combination of words that carry a 
specific scientific meaning (concepts) from each manuscript. The set of concepts
to some extent represent the content of the paper, both with respect to the subject of research 
and methods applied. To this end, we will use a ScienceWISE.info platform, specially tailored for such tasks.

The ScienceWISE.info platform \cite{sciencewise,martini2016sciencewise}
has been built to support the daily activities of research scientists.
The goal of the platform is to ``understand'' the interests of its users and to recommend them relevant 
newly submitted manuscripts. For this purpose, \texttt{arXiv} serves as one of the data source of new submissions.
In order to understand the research interests of the users, the platform extracts scientific concepts from the 
texts of the manuscripts and compares the concept vector of the manuscript and the corresponding concept vector of 
the user's research interest. Concept extraction approach implemented into this platform has two phases: 
i) automatic key phrase (concept candidate) extraction and
ii) [optional] crowd-sourced validations of scientific concepts.
During the first phase, each manuscript is scanned by the KPEX algorithm \cite{constantin2014automatic}.
The algorithm extracts key phrases from the text of the manuscript, and these key phrases serve as concept candidates.
Then, during the second step, the concept candidates are reviewed by the registered users of the platform who are permitted to validate the concepts.
The described procedure arrived at approximately 20,000 concepts as of the date when this paper was written.
About 500 of them have been marked as generic concepts assuming their generic meaning (the ones like \texttt{Energy}, 
\texttt{Mass} or \texttt{Temperature}).

Navigating over ScienceWISE.info platform at the end of September of 2019,
we accessed a collection with near 1.2M \texttt{arXiv} manuscripts with metadata and concepts 
list for each one (from April of 2007 till September of 2019).
As data is publicly available (anyone with access to the internet could get it), we scraped it to storage 
on our side with a convenient structure for further manipulations.
A detailed data parsing approach could be found at a GitHub repository \cite{Serhij_hub}.
For each manuscript, a set of concepts found within its text has been recorded.
The total number of unique extracted concepts is 19,446 and the number of concepts per manuscript 
varies in range 0 -- 1164. A similar dataset (it can be considered as a small subset of the described above) 
of 36386 articles in Physics domain 
have been previously investigated in 
\cite{palchykov2016ground,palchykov2018bipartite,palchykov2019modeling}.
Once the dataset is downloaded and prepared for the analysis, the first step of our investigation is to analyze 
the topological properties of the resulting concept network using the tools of Complex network theory as descibed
in the next Section \ref{III}.

\section{Concept networks and their topological features} \label{III}

\begin{figure}[!ht]
    \centering
    \includegraphics[width=0.4\textwidth]{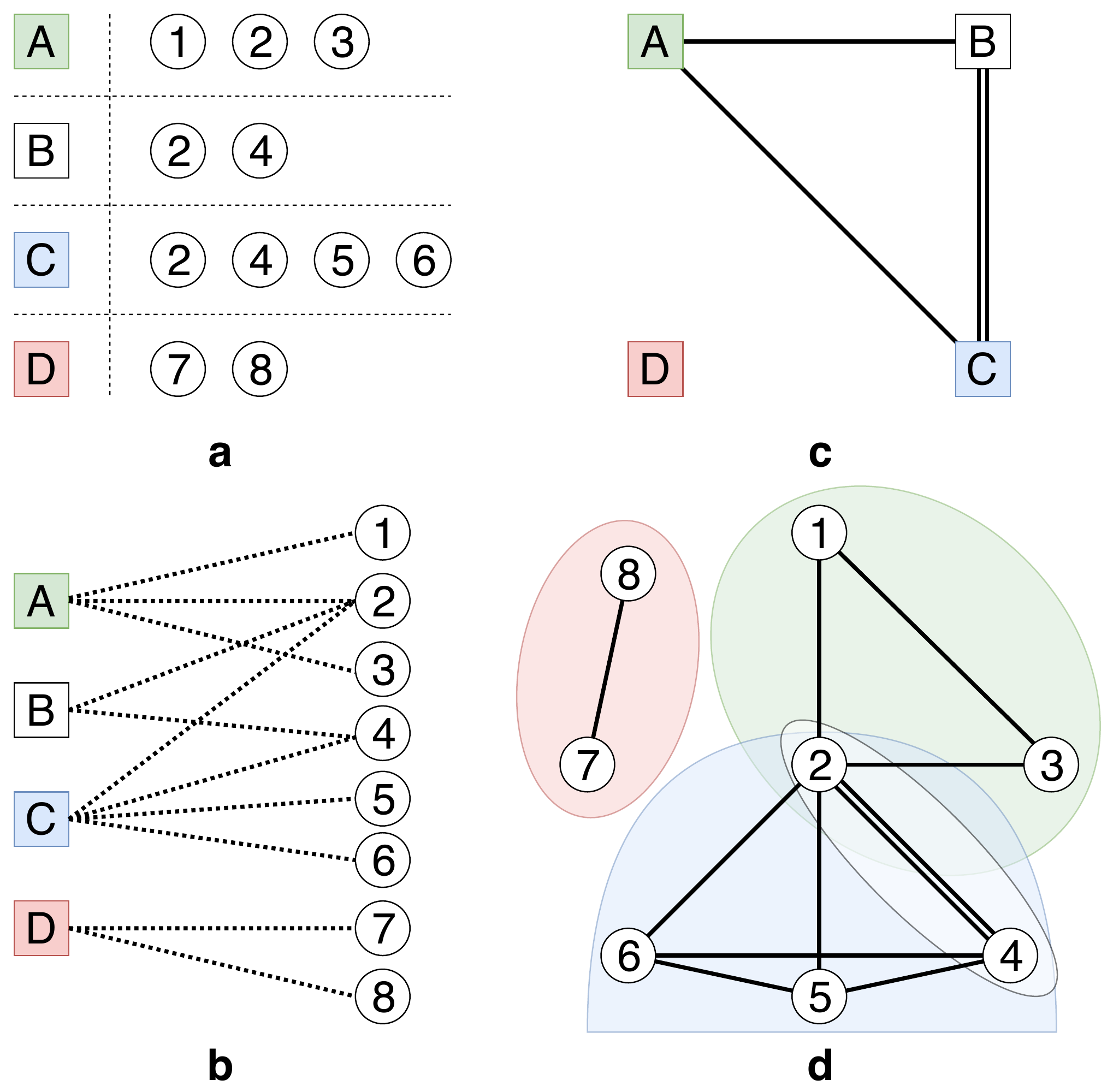}
    \caption{Illustration of the dataset and three network representations constructed from it.
    Squared nodes represent manuscripts and circles represent concepts.
    Panel {\bf a} illustrates a dataset of four manuscripts and the concepts identified within each of them.
    Panel {\bf b} is a bipartite network representation of the dataset.
    Panels {\bf c} and {\bf d} illustrate single-mode projections of the bipartite network to the manuscript
    and concept spaces, correspondingly. Connections between nodes represent how 
    many concepts has a given pair of manuscripts in common (panel {\bf c}) or how many manuscripts shared a 
    given pair of concepts (panel {\bf d}).
    }
    \label{fig1}
\end{figure}

The above described dataset may be naturally represented as a bipartite network, details of
network construction are shown in Fig. \ref{fig1}. Our further analysis is based on a single-mode
projection  of this network to the concept space, Fig. \ref{fig1} {\bf d}. We will call the
resulting network a concept network. There, a link between two nodes means that the corresponding 
concepts have appeared together in the lists of concepts for at least one manuscript.

To proceed with the analysis, we will consider two slices of data: the manuscripts submitted during the years
2013 and 2015. This will allow us, in particular, to compare some properties of a concept 
network as they evolve in time.
The first subset (the year 2013) network consists of 16,229 nodes, whereas for the year 2015 we arrived at 16,660 nodes.
We will refer to these networks as \texttt{g-2013} and \texttt{g-2015}, correspondingly. These networks share
15,431 concept-nodes in common.

To take into account link strength, we will implement two filtering procedures that keep only significant links in a network.
Within the first  procedure, we assign a weight  $w_{ij}$ to the link between sites $i$ and $j$ such that it equals the 
number of manuscripts in the dataset that contains concepts $i$ and $j$ simultaneously.
Then the simplest way to filter out insignificant links is to consider the hard threshold on the link weight.
Below we will consider threshold value $\omega = 10$ and keep the links for which $w_{ij} > \omega$. In our case, 
approximately $16.5\%$ of total links remain after such filtering.
If such procedure removes all links from a node, the node is removed too, so there are no isolated nodes in the network.
We will refer to the resulting networks as \texttt{w-2013} and \texttt{w-2015}, for the corresponding years.

A more sophisticated approach for filtering insignificant links is to consider the disparity filter proposed in Ref. \cite{serrano2009extracting}.
The key idea of the method is to calculate the probability $\alpha_{ij}$ that a given link is as strong or even stronger as observed 
in a random setting. This probability, known as $p$-value reads
\begin{equation} \label{eq:alpha}
    \alpha_{ij} = 1 - (k_i - 1) \int_0^{p_{ij}} (1 - x)^{k_i - 2} dx
\end{equation}
where $k_i$ is degree (i.e. the number of links) of the $i$-th node, and $p_{ij} = w_{ij}/(\sum_{j} w_{ij})$ is the normalized link weight.
Then one may set a threshold for $p$-value, and only the links with small enough $p$-value are kept, 
meaning that a random process can not arrive at a link with such weight.
In our analysis we set a threshold for $p$-value $\rho = 0.1$, and keep a link between $i$ and $j$ if $\alpha_{ij} < \rho$ or 
$\alpha_{ji} < \rho$, i. e. if it is significant from at least one node standpoint.
As a result of such procedure, approximately $85\%$ of links are removed as insignificant.
The resulting concept networks for the corresponding years will be referred to as \texttt{d-2013} and \texttt{d-2015}.

With the networks at hand, it is straightforward to compare them measuring standard indices that quantify their
different features. In Table \ref{tab:network_characteristics} we report some values obtained by us, 
a more comprehensive comparison can be found in Ref. \cite{Serhij_diploma}. There,  besides
the number of network nodes $N$ and links $L$  and mean and maximal node degrees
 $\langle{k}\rangle$,  $k_{\rm max}$,
we provide the values, that characterize
network size (mean and maximal shortest path lengths  $l$, $l_{\rm max}$ measured as a shortest number of steps between two different nodes)
and correlations in network structure. To quantify correlations, we measured  mean clustering coefficient $\langle{c}\rangle$,
global transitivity $C$ and assortativity $r$. The clustering coefficient $c_i$ of node $i$ describes the 
level of connectivity among its neighbours: $c_i = \frac{2 m_i}{k_i (k_i - 1)}$ where $m_i$ is the number of existing connections 
among $k_i$ neighbouring nodes. Therefore, the mean value $\langle{c}\rangle$ of $c_i$, averaged over all nodes in the network, 
characterizes the local density of neighborhood links in the entire network.
Instead of calculating the average value of local measurements, global transitivity
$C$ is defined as a ratio between the total number of connected triplets in the network and the number of all 
possible triangles. In turn, assortativity $r$ is defined as 
Pearson correlation coefficient between node degrees on both ends over existing link 
\cite{newman2003structure,holovatch2006complex,newman2010networks}.

\begin{table*}[htbp]
    \caption{Aggregated characteristics of concepts networks.
    $N$,  $L$: number of nodes and links; $\rho = 2L / N(N-1)$: density of links; $\langle{k}\rangle$,  $k_{\rm max}$: mean and maximal node degree;
    $l$, $l_{\rm max}$:  mean and maximal shortest path lengths; $\langle{c}\rangle$: mean clustering coefficient; $C$:
    global transitivity; $r$: assortativity. See the text for more description.
    }
    \centering
    \begin{tabular}{l|r|r|r|r|r|r|r|r|r|r}
    \hline\hline
    network & $N$ & $L, \times 10^6$ & $\rho, \%$ & $\langle{k}\rangle$ & $k_{\rm max}$ & $l$ & $l_{\rm max}$ & $\langle{c}\rangle$ & $C$ & $r$\\ \hline
    \texttt{g-2013} & 16,229 & 11.1 & 8.46 & 1,373 & 15,345 & 1.92 & 3 & 0.77 & 0.37 & -0.324\\
    \texttt{g-2015} & 16,660 & 12.7 & 9.12 & 1,520 & 15,935 & 1.91 & 4 & 0.77 & 0.38 & -0.325\\\hline
    \texttt{w-2013} & 9,999 & 1.8 & 3.69 & 369 & 8,856 & 2.00 & 4 & 0.89 & 0.28 & -0.390\\
    \texttt{w-2015} & 10,770 & 2.2 & 3.84 & 414 & 9,661 & 2.00 & 4 & 0.89 & 0.28 & -0.382\\\hline
    \texttt{d-2013} & 13,358 & 1.6 & 1.84 & 246 & 11,665 & 2.01 & 4 & 0.90 & 0.14 & -0.375\\
    \texttt{d-2015} & 13,969 & 1.9 & 1.92 & 268 & 12,367 & 2.00 & 5 & 0.89 & 0.14 & -0.368\\
    \hline\hline
    \end{tabular}
    \label{tab:network_characteristics}
\end{table*}

Our analysis of the topology of the concept network indicates that observed concept networks are heterogeneous graphs that obey internal 
clustering (community structure) and hierarchical organization. These properties of a concept network are independent of the subset 
of data used (constructed from 2013 and 2015 year data). These features, however, are more pronounced once weak links have been removed.
As it follows from the comparison of data obtained for different years and via different procedures of relevant link determination, 
cf. Table~\ref{tab:network_characteristics}, complex networks under consideration attain a range of universal features that do not change 
with time and characterize the system of concepts as a whole.
In particular, they are the small world networks 
\cite{albert2002statistical,dorogovtsev2013evolution,newman2003structure,holovatch2006complex,barrat2008dynamical,newman2010networks}
characterized by a small size (mean shortest path and maximal shortest path values) and large value of clustering coefficient.
The last also brings about the presence of strong correlations.
Moreover, an essential difference between the clustering coefficient and global transitivity serves as evidence of possible community structure.
In turn, the negative value of assortativity suggests that they are disassortative networks where a group of central nodes (hubs) serves as 
common attraction points for nodes with lower degree values.

\section{Scientific innovations and concept embedding} \label{IV}

In this section, we 
investigate the possibility to detect in advance fields where scientific innovations may emerge.
In particular, we are interested in the questions of the prediction power of concept embedding.

Investigation of scientific innovation emergence is not straightforward.
The simplification adopted in frames of this paper considers innovations as the appearance of a new statistically significant link between nodes that 
previously were not linked to each other.
In this way, the emergence of such a link is treated as a novelty introduced into the graph of scientific concepts.

We proceed by considering a network of scientific concepts built upon manuscripts submitted to \texttt{arXiv} during the year 2013.
Let us consider a pair of concepts $i$ and $j$.
In terms of link existence, these concepts may be either connected by a link or disconnected, meaning no link between $i$ and $j$.
The fraction of pairs connected by links equals to the density of links, $\rho = 2L / N(N-1)$, in the corresponding concept network. 
For a \texttt{g-2013} network  it has a value $\rho = 8.46\%$, see Table \ref{tab:network_characteristics}.
Some of the links that carry low weight may be considered as spurious links rather than statistically significant, meaning that they could arise as a 
result of noise rather than a real coupling between the corresponding concepts.
In this paper, we consider two alternative ways to filter out such spurious links: i) naive filtering by setting up a link weight threshold and 
ii) disparity filtering that employs statistical significance testing,
as explained in Section \ref{III}.

With the thresholds set above ($\omega=10$ and $\rho=0.1$), majority of the concept pairs (out of about 130M potential connections) are either 
disconnected or are connected by spurious links.
Namely, $98.6\%$ of concept pairs out of all possible $N(N-1)/2$ pairs are disconnected or connected by weak links ($\omega_{ij} \leq \omega$, referred below as 
\texttt{weak/missing links}) and $98.8\%$ of pairs are either disconnected or connected by a statistically insignificant links 
($\alpha_{ij}\geq\rho$, referred below as statistically \texttt{insignificant links}).

Some of these pairs may become connected in the future by strong or statistically significant connections.
The emergence of such connections is referred in this paper as scientific innovations.
Our analysis indicates that only 564,330 pairs ($0.43\%$) became strongly connected ($w_{ij} > 10$) in 2015 out of 129,837,653 weakly 
connected/disconnected pairs in 2013.
Disparity filter arrives at a similar picture.
Only 475,788 pairs ($0.37\%$) became statistically significant in 2015 out of 130,039,148 insignificant/disconnected pairs in 2013.
To conclude, less than $0.5\%$ of \texttt{weak/missing links} or statistically \texttt{insignificant links} between concepts in 2013 became 
strong/significant in the year 2015.

Thus the questions of our interest are related to forecasting the pairs where such innovations may emerge given the number (fraction) 
of such connections is known.
In particular, we are interested in the power of concept embedding technique \cite{mikolov2013distributed,lerer2019pytorch}
to distinguish between the pairs of concepts that will become connected vs the pairs that will stay disconnected in the future.
The key assumptions are that i) concepts that appear in a similar context will have close enough vectors in embedded space and ii) that 
the concepts that carry similar content are more likely to become connected in the future.

For this reason we use concept co-occurrence matrix for year 2013 and embedded each concept vector in 100 dimensional space using 
PyTorch-BigGraph \cite{lerer2019pytorch}.
The whole detailed pipeline we used for described graphs and embeddings formulation can be found at the GitHub repository
\cite{Serhij_hub}.
As a result, each concept $i$ becomes associated with a vector $\vec{v_i}$ in the embedded space.
The similarity $s_{ij}$ between a pair of concepts $i$ and $j$ is then calculated as a cosine similarity between the corresponding vectors 
$\vec{v_i}$ and $\vec{v_j}$.

Once similarities $s_{ij}$ between concept vectors in embedded space have been calculated for each pair of concepts $i$ and $j$, we divide 
all pairs of concepts into two groups:
i) \texttt{Strong embedding similarity group} and ii) \texttt{Weak embedding similarity group}.
To distribute pairs of nodes/concepts among the groups, we put an arbitrarily selected threshold of $\zeta = 0.6$. 
The pairs of concepts for which embedding similarity $s_{ij} \leq \zeta$ are assigned to \texttt{Weak embedding similarity group}, 
for convenience, we will refer to the corresponding pairs as \texttt{dissimilar concepts}.
Instead, if the embedding similarity between concepts $i$ and $j$, $s_{ij} > \zeta$, the corresponding pair is assigned to a 
\texttt{Strong embedding similarity group} and will be referred below as \texttt{similar concepts}.
We expect that the selection of the other value of $\zeta$ threshold will not change the qualitative results of our analysis.
Especially, because pairs on both extremes of embedding similarity will eventually be assigned to different groups.

The results of our analysis indicate significant differences in the allocation of pairs of concepts among embedding similarity 
groups for weakly and strongly connected pairs of nodes in the network.
While only $1.2\%$ of \texttt{weak/missing links} in \texttt{g-2013} falls into \texttt{similar concepts} 
group, this fraction is much higher for \texttt{strong links}, reaching $22.8\%$.
Similar results have been observed if one uses a disparity filter instead of link weight threshold filter.
Thus, we expect that the grouping of pairs of nodes using embedding similarity improves predictions of
the pairs of concepts where statistically significant links will be established in the future.

With the data about the concept network for the year 2013 at hand, let us now consider the network of scientific concepts constructed 
from manuscripts submitted to \texttt{arXiv} during the year 2015.
Below we perform preliminary analysis rather than propose a predictive model.

Comparing the networks constructed from data of years 2013 and 2015, we see that the majority of strongly connected concept pairs in 2015 
were connected by strong links in 2013 too.
Table~\ref{tab:table_percentage_links_2015_weight} shows that about $90\%$ of strong links in 2013 remained strong in the year 2015.
\begin{table}[htbp]
    \caption{Percentage of concept pairs that belong to a specific combination of link weight group and embedding similarity group in 
    2013 that either remained or became strong in 2015.}
    \centering
    \begin{tabular}{p{2.1cm}|p{2cm}|p{2cm}}
    & \texttt{dissimilar concepts} during 2013 & \texttt{similar concepts} during 2013 \\ \hline
    \texttt{weak links} during 2013 & $0.4\%$ & $3.19\%$ \\ \hline
    \texttt{strong links} during 2013 & $88.89\%$ & $94.34\%$ \\ \hline
    \end{tabular}
    \label{tab:table_percentage_links_2015_weight}
\end{table}
If we take into account grouping by concept embedding similarity, we observe additional segregation: strong links with low embedding 
similarity in the year 2013 remained strong in the year 2015 in almost $89\%$ of cases,
while strong links with strong embedding similarity in the year 2013 remained strong in the year 2015 for more than $94\%$ of cases.
These results lead us to the following conclusions.
First, if a link between two concept-nodes exists and this is a strong link, then it is likely that the link will exist in the future, 
and it will remain the strong one.
In other words, the strength of a link is a good predictor for a link to belong to the same category in the future.
Second, strong links with high concept embedding similarity have higher chances to remain strong in the future than strong links that 
are characterized by low embedding similarity.

On the other side, weak links evolve to strong links quite rarely.
Only $0.4\%$ of weak links in 2013 evolved to strong links in year 2015.
However, classification of concepts pairs by their embedding similarity allowed us to identify a subgroup of these pairs for which the 
probability of becoming strong connections raises to $3.19\%$, i. e. in about 8 times.
Even though the concept embedding similarity does not point the ``future'' emergence of a new strong link in the network exactly, the results 
of our analysis indicate its power as one of the features to be used in such predictions.

Similar results have been obtained if we use classification of links between pairs of concepts using statistical significance testing instead 
of link weight threshold, see Table~\ref{tab:table_percentage_links_2015_disparity}.

\begin{table}[htbp]
    \caption{Percentage of concept pairs that belong to a specific combination of link significance group and embedding similarity group in 
    2013 that either remained or became strong in 2015.}
    \centering
    \begin{tabular}{p{2.5cm}|p{2cm}|p{2cm}}
    & \texttt{dissimilar concepts} during 2013 & \texttt{similar concepts} during 2013 \\ \hline
    \texttt{insignificant links} during 2013 & $0.3\%$ & $3.01\%$ \\ \hline
    \texttt{significant links} during 2013 & $82.71\%$ & $91.06\%$ \\ \hline
    \end{tabular}
    \label{tab:table_percentage_links_2015_disparity}
\end{table}

Thus, independent of the method used to classify pairs of concepts, either using link weight threshold or statistical significance testing,
the results of our analysis indicate the ability of concept embedding similarity in predicting scientific innovations, i. e. the emergence of strong 
or statistically significant links in a concept network.

\section{Conclusions and Outlook} \label{V}

The goal of our work was to analyze the possibilities of innovation emergence in the course of knowledge generation.
To this end, we have investigated the structure and dynamics of connections between scientific concepts that constitute a body of research 
papers, as recorded in the \texttt{arXiv} repository \cite{arxiv}.
We have applied two methods, concept embedding and network analysis, to quantify properties of sets of concepts and to predict the emergence 
of new links (innovations) between different concepts.
We have shown that whereas each of the above methods is a powerful tool to define certain features of a system of concepts,
it is the combination of these two methods that leads to a synergetic effect and allows to forecast dynamics of new links creation and evolution 
of a system as a whole.
The main results obtained in the course of our analysis include the following:

\begin{itemize}
    \item We have represented a system of concepts of scientific papers in the form of a complex network.
    Different nodes in this network correspond to different concepts, and a link between two nodes-concepts means that they were exploited 
    in the same paper.
    We have determined the quantitative characteristics of a complex network of concepts and their evolution with time, and the data 
    is given in Table~\ref{tab:network_characteristics}.
    
    \item We have used two complementary approaches to define the presence of a strong link between two nodes, i. e. of a link that serves 
    as evidence of a relevant connection.
    In one approach, the criterion is given by a link weight.
    The second method takes into account subtle information about network intrinsic structure \cite{serrano2009extracting}.
    Corresponding data is shown in Table~\ref{tab:network_characteristics}.
    
    \item As is follows from the comparison of data obtained for different years and via different procedures of relevant link determination, 
    see Table~\ref{tab:network_characteristics},
    complex networks under consideration attain a range of universal features that do not change with time and characterize the system 
    of concepts as a whole.
    In particular, they are the small world networks characterized by small size (mean the shortest path and maximal shortest path values) 
    and large value of the clustering coefficient.
    The last also brings about the presence of strong correlations.
    Moreover, an essential difference between the clustering coefficient and global transitivity serves as evidence of possible community structure.
    In turn, the negative value of assortativity suggests that they are disassortative networks where a group of central nodes (hubs) serves as 
    common attraction points for nodes with lower degree value.
    
    \item Concept embedding technique enabled us to find out proximity (by context, by subject, or related in any other way) between different concepts.
    With a measure of proximity at hand, we were in a position to compare it with the dynamics of new links emergence between different concepts.
    In turn, this enables one to reveal groups of concepts (subsequently – fields of knowledge) where innovations are probable to emerge.
    Corresponding statistical analysis is summarized in Tables~\ref{tab:table_percentage_links_2015_weight} and 
    \ref{tab:table_percentage_links_2015_disparity}.
\end{itemize}

The results obtained in this study may be useful both from the fundamental point of view, contributing to our understanding of how the knowledge 
is formed, as well as they may have the practical implementation.
In particular, the methodology elaborated in the course of our analysis can be used to detect fields where innovations have a higher probability 
of appearing.
A natural way to continue the analysis presented here is to evaluate practical outcomes (i. e. impact) of papers, where the higher probability 
of innovation is predicted.
With the scientometric data at hand, such a task is not much time consuming and will be a subject of future work.
Another work in progress is to suggest a model predicting the emergence of statistically significant links between already existing concepts.

{\bf Acknowledgement.} A part of this work has been performed in the frames of the Master thesis by S.B. in Ukrainian Catholic
University (Lviv, Ukraine).



\begin{thebibliography}{10}
\providecommand{\url}[1]{#1}
\csname url@samestyle\endcsname
\providecommand{\newblock}{\relax}
\providecommand{\bibinfo}[2]{#2}
\providecommand{\BIBentrySTDinterwordspacing}{\spaceskip=0pt\relax}
\providecommand{\BIBentryALTinterwordstretchfactor}{4}
\providecommand{\BIBentryALTinterwordspacing}{\spaceskip=\fontdimen2\font plus
\BIBentryALTinterwordstretchfactor\fontdimen3\font minus
  \fontdimen4\font\relax}
\providecommand{\BIBforeignlanguage}[2]{{%
\expandafter\ifx\csname l@#1\endcsname\relax
\typeout{** WARNING: IEEEtran.bst: No hyphenation pattern has been}%
\typeout{** loaded for the language `#1'. Using the pattern for}%
\typeout{** the default language instead.}%
\else
\language=\csname l@#1\endcsname
\fi
#2}}
\providecommand{\BIBdecl}{\relax}
\BIBdecl

\bibitem{zhmud2008origin}
L.~Zhmud, \emph{The origin of the History of Science in Classical
  Antiquity}.\hskip 1em plus 0.5em minus 0.4em\relax Walter de Gruyter, 2008,
  vol.~19.

\bibitem{lewens2016meaning}
T.~Lewens, \emph{The meaning of science: An introduction to the philosophy of
  science}.\hskip 1em plus 0.5em minus 0.4em\relax Hachette UK, 2016.

\bibitem{mryglod2016quantifying}
O.~Mryglod, Yu.~Holovatch, R.~Kenna, and B.~Berche, ``Quantifying the evolution
  of a scientific topic: reaction of the academic community to the chornobyl
  disaster,'' \emph{Scientometrics}, vol. 106, no.~3, pp. 1151--1166, 2016.
  

\bibitem{Krenn20}
M.~Krenn and A.~Zeilinger, ``Predicting research trends with semantic and neural networks
with an application in quantum physics,''
\emph{arXiv preprint arXiv:1906.06843v2}, 2020.  

\bibitem{leydesdorff2015scientometrics}
L.~Leydesdorff and S.~Milojevi{\'c}, ``Scientometrics'' in:  Lynch Micheal,
  editor. \emph{International Encyclopedia of Social and Behavioral Sciences}, 2015.

\bibitem{berche2016academic}
B.~Berche, Yu.~Holovatch, R.~Kenna, and O.~Mryglod, ``Academic research groups:
  evaluation of their quality and quality of their evaluation,'' in
  \emph{J. Phys.: Conf. Ser.}, vol. 681, 012004, 2016.

\bibitem{Thurner19}
S.~Thurner, W.~Liu, P.~Klimek, and S.~A.~Cheong, ``The role of mainstreamness 
and interdisciplinarity for the relevance of scientific papers,''
\emph{arXiv preprint arXiv:1910.03628}, 2019.

  
\bibitem{zeng2017science}
A.~Zeng, Z.~Shen, J.~Zhou, J.~Wu, Y.~Fan, Y.~Wang, and H.~E. Stanley, ``The
  science of science: From the perspective of complex systems,'' \emph{Physics
  Reports}, vol. 714, pp. 1--73, 2017.

\bibitem{iacopini2018network}
I.~Iacopini, S.~Milojevi{\'c}, and V.~Latora, ``Network dynamics of innovation
  processes,'' \emph{Phys. Rev. Lett.}, vol. 120, no.~4, p. 048301,
  2018.

\bibitem{uzzi2013atypical}
B.~Uzzi, S.~Mukherjee, M.~Stringer, and B.~Jones, ``Atypical combinations and
  scientific impact,'' \emph{Science}, vol. 342, no. 6157, pp. 468--472, 2013.
 

\bibitem{arxiv}
arXiv gives an open access to 1,640,097 e-prints in Physics,
  Mathematics, Computer Science, Quantitative Biology, Quantitative Finance,
  Statistics, Electrical Engineering and Systems Science, and Economics,
  accessed: 2020-01-03. [Online]. Available: \url{https://arXiv.org}

\bibitem{sciencewise}
The ScienceWISE project aims to develop a scientist-generated
  on-line knowledge base fully integrated into the physics ArXiv.org, accessed:
  2020-01-06. [Online]. Available: \url{http://ScienceWISE.info}

  
\bibitem{holovatch2006complex}
Yu.~Holovatch, O.~Olemskoi, C.~von~Ferber, T.~Holovatch, O.~Mryglod,
  I.~Olemskoi, and V.~Palchykov, ``Complex networks,'' \emph{Journ. Phys. Stud.}, 
  vol.~10, pp. 247--289, 2006.

\bibitem{newman2010networks}
\BIBentryALTinterwordspacing
M.~Newman, \emph{Networks: An Introduction}.\hskip 1em plus 0.5em minus
  0.4em\relax OUP Oxford, 2010. [Online]. Available:
  \url{https://books.google.com.ua/books?id=LrFaU4XCsUoC}
\BIBentrySTDinterwordspacing

\bibitem{barabasi2016network}
A.-L. Barab{\'a}si \emph{et~al.}, \emph{Network science}.\hskip 1em plus 0.5em
  minus 0.4em\relax Cambridge university press, 2016.

\bibitem{holovatch2018statistical}
Yu.~Holovatch, M.~Dudka, V.~Blavatska, V.~Palchykov, M.~Krasnytska, and
  O.~Mryglod, ``Statistical physics of complex systems in the world and in
  Lviv,'' \emph{Journ. Phys. Stud.}, vol.~22, no. 2801, p.~21,
  2018.

\bibitem{mikolov2013distributed}
T.~Mikolov, I.~Sutskever, K.~Chen, G.~S. Corrado, and J.~Dean, ``Distributed
  representations of words and phrases and their compositionality,'' in
  \emph{Advances in Neural Information Processing Systems}, 2013, pp.
  3111--3119.

\bibitem{lerer2019pytorch}
A.~Lerer, L.~Wu, J.~Shen, T.~Lacroix, L.~Wehrstedt, A.~Bose, and
  A.~Peysakhovich, ``Pytorch-biggraph: A large-scale graph embedding system,''
  \emph{arXiv preprint arXiv:1903.12287}, 2019.

\bibitem{martini2016sciencewise}
A.~Martini, A.~Lutov, V.~Gemmetto, A.~Magalich, A.~Cardillo, A.~Constantin,
  V.~Palchykov, M.~Khayati, P.~Cudr{\'e}-Mauroux, A.~Boyarsky \emph{et~al.},
  ``Sciencewise: Topic modeling over scientific literature networks,''
  \emph{arXiv preprint arXiv:1612.07636}, 2016.

\bibitem{constantin2014automatic}
A.~Constantin, ``Automatic structure and keyphrase analysis of scientific
  publications,'' Ph.D. dissertation, The University of Manchester (United
  Kingdom), 2014.
  
\bibitem{Serhij_hub} github.com/sergibro/concept-graphs

\bibitem{palchykov2016ground}
V.~Palchykov, V.~Gemmetto, A.~Boyarsky, and D.~Garlaschelli, ``Ground truth?
  Concept-based communities versus the external classification of physics 
  manuscripts,'' \emph{EPJ Data Science}, vol.~5, no.~1, p.~28, 2016.

\bibitem{palchykov2018bipartite}
V.~Palchykov and Yu.~Holovatch, ``Bipartite graph analysis as an alternative to
  reveal clusterization in complex systems,'' in \emph{2018 IEEE Second
  International Conference on Data Stream Mining \& Processing (DSMP)}.\hskip
  1em plus 0.5em minus 0.4em\relax IEEE, 2018, pp. 84--87.

\bibitem{palchykov2019modeling}
V.~Palchykov and Yu.~Holovatch, \emph{to be published}, 2020.

\bibitem{serrano2009extracting}
M.~{\'A}. Serrano, M.~Bogun{\'a}, and A.~Vespignani, ``Extracting the
  multiscale backbone of complex weighted networks,'' \emph{Proc.Nat.
  Acad. Sci.}, vol. 106, no.~16, pp. 6483--6488, 2009.

\bibitem{Serhij_diploma}  
S.~Brodiuk.   \emph{Concept embedding and network analysis
of scientific innovations emergence}.\hskip 1em plus 0.5em minus 0.4em\relax Master thesis,
Ukrainian Catholic University, Lviv, 2019.

\bibitem{newman2003structure}
M.~E. Newman, ``The structure and function of complex networks,'' \emph{SIAM
  Review}, vol.~45, no.~2, pp. 167--256, 2003.

\bibitem{albert2002statistical}
R.~Albert and A.-L. Barab{\'a}si, ``Statistical mechanics of complex
  networks,'' \emph{Rev. Mod. Phys.}, vol.~74, no.~1, p.~47, 2002.

\bibitem{dorogovtsev2013evolution}
S.~N. Dorogovtsev and J.~F. Mendes, \emph{Evolution of networks: From
  biological nets to the Internet and WWW}.\hskip 1em plus 0.5em minus
  0.4em\relax OUP Oxford, 2013.

\bibitem{barrat2008dynamical}
A.~Barrat, M.~Barthelemy, and A.~Vespignani, \emph{Dynamical processes on
  complex networks}.\hskip 1em plus 0.5em minus 0.4em\relax Cambridge
  university press, 2008.

\end{thebibliography}
\end{document}